\documentclass[pra,twocolumn,amsmath,amssymb,superscriptaddress,longbibliography]{revtex4-1}

\usepackage{graphicx,upgreek,color,mathtools,bm,mathptmx,wasysym,mathptmx}
\usepackage[hyphenbreaks]{breakurl}
\usepackage[colorlinks=true,linkcolor=blue,citecolor=blue,urlcolor =blue]{hyperref}
\DeclareMathOperator{\Var}{Var}

\begin{document}

\title{Intensity-Based Axial Localization at the Quantum Limit}

\author{J. \v{R}eh\'a\v{c}ek} 
\affiliation{Department of Optics,
 Palack\'y University, 17. listopadu 12, 771 46 Olomouc, 
Czech Republic}

\author{M. Pa\'{u}r} 
\affiliation{Department of Optics,
 Palack\'y University, 17. listopadu 12, 771 46 Olomouc, 
Czech Republic}

\author{B. Stoklasa}
\affiliation{Department of Optics,
 Palack\'y University, 17. listopadu 12, 771 46 Olomouc, 
Czech Republic}

\author{D. Koutn\'y}
\affiliation{Department of Optics,
 Palack\'y University, 17. listopadu 12, 771 46 Olomouc, 
Czech Republic}

\author{Z. Hradil}
\affiliation{Department of Optics,
 Palack\'y University, 17. listopadu 12, 771 46 Olomouc, 
Czech Republic}

\author{L. L. S\'{a}nchez-Soto} 
\affiliation{Departamento de \'Optica,
Facultad de F\'{\i}sica, Universidad Complutense, 
28040~Madrid,  Spain} 
\affiliation{Max-Planck-Institut f\"ur die Physik des Lichts,
Staudtstra{\ss}e 2, 91058 Erlangen, Germany}

\begin{abstract}
  We derive fundamental precision bounds for single-point axial
  localization. For the case of a Gaussian beam, this ultimate limit
  can be achieved with a single intensity scan, provided the camera is
  placed at one of two optimal transverse detection planes. Hence, for
  axial localization there is no need of more complicated detection
  schemes. The theory is verified with an experimental demonstration
  of axial resolution three orders of magnitude below the classical
  depth of focus.
\end{abstract}

\maketitle
  
\emph{Introduction.---}
The maximum spatial resolution attainable in classical microscopy 
is usually established in terms of the Abbe-Rayleigh
criterion~\cite{Rayleigh:1879aa,Abbe:1873aa}. However, it is notorious
that this criterion is based on heuristic notions and is an inadequate
performance measure for current quantitative imaging~\cite{Ram:2006aa}.

Indeed, several modern techniques, gathered under the broad
denomination of superresolution
microscopy~\cite{Hell:2007aa,Huang:2009aa,
  Schermelleh:2010aa,Leung:2011aa,Huszka:2019aa}, are capable of
achieving a striking increase in resolution by more than one order of
magnitude in comparison with the length scale set by the Abbe-Rayleigh
criterion.  An important class of these techniques (which includes,
among others, stimulated-emission-depletion
microscopy~\cite{Hell:1994ab}, photoactivated-localization
microscopy~\cite{Betzig:2006aa}, PSF engineering~\cite{Huang:2008aa,
  Pavani:2009aa,Jia:2014aa,Tamburini:2006aa,Paur:2018aa}, and
multiplane detection~\cite{Juette:2008aa,Dalgarno:2010aa,
  Abrahamsson:2012aa}) relies on a very accurate localization of point sources.

For three-dimensional imaging, extracting the emitter axial position
is an enduring challenge that has been extensively
investigated~\cite{Diezmann:2017aa}. Yet, finding the optimal
depth precision attainable by any such microscope
engineering approach has been only recently 
tackled~\cite{Tsang:2015aa,Backlund:2018aa}. The basic idea  is to
use the quantum Fisher information (qFI) and the associated
Cram\'er-Rao bound (qCRB) to get a measurement-independent
limit~\cite{Petz:2011aa}, much in the same vein as Tsang and coworkers
did to quantify two-point resolution~\cite{Tsang:2016aa, Nair:2016aa,
  Ang:2016aa,Tsang:2017aa}.

In this Letter, we address this fundamental question from a different
perspective. By identifying the unitary transformation that embodies
the action of the system and its corresponding generator, we get in a
very transparent way the ensuing qCRB. More important, we do find the
optimal measurement reaching such a  limit.

We focus here on direct imaging, for this is the simplest method
available in the laboratory. Of course, one could rightly argue that
in this way all the phase information is wasted. Surprisingly, we
demonstrate that direct detection can saturate the quantum limits with
a single intensity scan, as long as the camera is placed in one 
optimal transverse detection plane. This might be of utmost
importance for any application demanding extreme stringent
localization, as it only requires very simple and feasible equipment.

\emph{Theoretical model.---}
To simplify the details as much as possible, we take the
waist of a focused beam as our object. The task is to estimate the distance from
this object to a detection plane. In the following, we use the Dirac
notation to represent the field, as it allows to extend the theory to
any type of light source.

If the beam in the object plane is represented  by the pure state
$|\Psi (0) \rangle$, the axial displacement is described  by a unitary operation
\begin{equation}
  \label{eq:psiz}
|\Psi (z)\rangle = e^{i G \,  z} \; | \Psi (0) \rangle \, ,
\end{equation}
the Hermitian operator $G$ being corresponding generator. To identify the action
of $G$ in a more precise way, it is convenient to use the
transverse-position representation  $\Psi (x,y; z) = \langle x,y |
\Psi (z) \rangle$. Given the form of Eq.~\eqref{eq:psiz}, we have that
\begin{equation}
  \partial_{z} \Psi (x,y; z) = i G \,  \Psi (x,y; z) \, ,
\end{equation}
which is consistent with the paraxial wave equation $ 2 i k
\partial_{z} \Psi (x,y; z) =  \nabla^{2}_{T}  \Psi (x,y; z)$  if
\begin{equation}
  \label{eq:gen}
  G \mapsto \frac{1}{2k} \nabla^{2}_{T} \, ,
  \end{equation}
where $k$ is the wavenumber and $\nabla^{2}_{T} = \partial_{xx} +
\partial_{yy}$ is the transverse Laplacian.

For a more tractable analysis and experiment, here we
assume a normalized Gaussian beam
\begin{equation}
  \label{eq:Gauss}
  \Psi(r; z)=\frac{2}{w(z)} e^{-\frac{r^2}{w^{2}(z)}}
  \exp \left ( -i \left [ k z + \frac{kr^2}{2 R(z)}- \phi(z) \right ]
  \right ) \,  ,
\end{equation}
although the results are largely independent of this choice. Notice that,
given the cylindrical symmetry, the beam depends exclusively on the
radial coordinate $r$.  The field distribution in Eq.~\eqref{eq:Gauss}  is
determined by  the beam waist $w_{0}$ and the Rayleigh range $z_R$
through  $w^{2}(z) = w_{0}^{2} [ 1+(z/z_R)^2]$, $R(z)=z[
1+(z_R/z)^{2}]$,  $\phi(z)= \arctan(z/z_R)$, and
$z_{R} =\pi w_{0}^{2}/\lambda$.

The detection plane is placed at $z$ and therein we perform a
measurement that we do not need to specify by the time being.  To
quantify the information about $z$ available in the measured 
signal we use  the qFI, which, for pure states, as it is our case, is
given by $\mathcal{Q}( {z} ) = 4 \Var (G)$, 
where $\Var$ is the variance computed in the initial state. Given the 
representation~\eqref{eq:gen} for $G$, a direct 
calculation shows that for the Gaussian beam one has
\begin{equation}
\label{eq:quant_lim}
\mathcal{Q} (z) = \frac{1}{z_R^{2} } \, ,
\end{equation}
which turns out to be constant.
The qCRB~\cite{Helstrom:1976ij,Holevo:2003fv} ensures then that  the
variance of any unbiased estimator $\widehat{z}$ of the displacement
$z$  is bounded by the reciprocal of the qFI; viz, $\Var ( \widehat{z}
) \ge  1/\mathcal{Q}(z)$.  In consequence, the
lower bound on axial-measurement errors (per single detection) is
precisely the Rayleigh range. This agrees with the result recently
found in Ref.~\cite{Zhou:2019aa}, which discusses the ultimate limits for
two-point axial resolution.

\emph{Direct detection.---}
In general, the qFI would be distributed between the phase and
intensity variations of the measured beam. One would naively expect
that intensity detection, discarding all phase information,
cannot saturate the quantum limit~\eqref{eq:quant_lim}. We will show 
that, contrary to this belief, this is not the case when the
detector is appropriately placed.

\begin{figure}
\includegraphics[width=0.95\columnwidth]{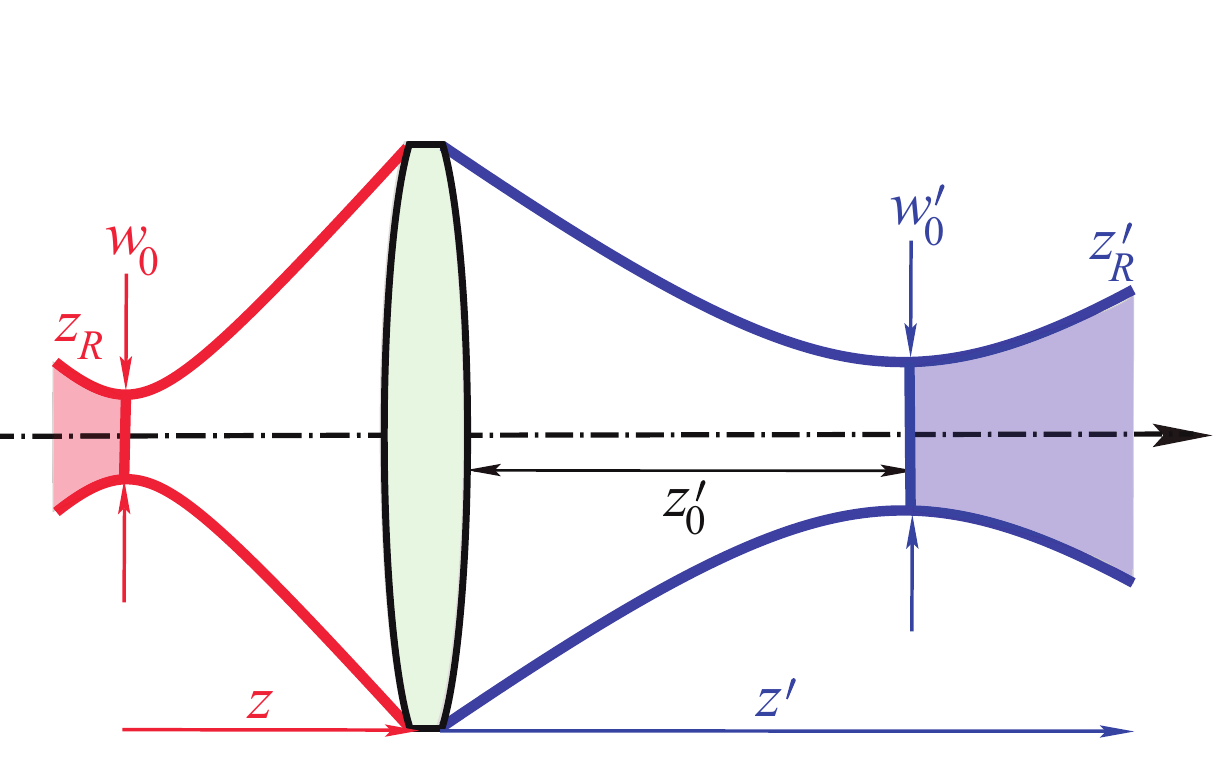}
\caption{Scheme of the axial localization experiment with a relay optical system.}
\label{figGauss}
\end{figure}

We model the light detection as a random process and, consequently, we
interpret the normalized beam intensity $p(r|z)=|\Psi(r;z)|^2$ as the
probability density of a single detection event  at $r$ conditional on
the value of $z$. We assume that detection is limited by shot noise,
which obeys a Poisson distribution~\cite{Fuhrmann:2004aa}. This
simplified approach ignores nonclassical effects, as bunching or
entanglement, but is nonetheless relevant to practical microscopy.
In addition, we ignore finite spatial extent and nonzero pixel
size. Under these hypothesis,  the classical Fisher information about
$z$  per single detection is   
\begin{equation}
\label{fishinf}
\mathcal{F} (z) =  2 \pi \int_{0}^{\infty} r \,  \frac{[ \partial_z
  p(r|z) ]^2}{p(r|z)} \, dr \, ,
\end{equation}
and the associated CRB quantifies the axial localization error for direct
detection.  For a Gaussian beam, $ p(r|z) =[ \pi w^2(z)/2]^{-1}  
\exp [- 2 r^2/w^2(z)]$, so that 
\begin{equation}
\label{res1}
\mathcal{F} (z) = \frac{\partial_z w^{2}(z)}{w^{2}(z)} = 
\frac{1}{4 R^{2}(z)} = \frac{1}{4 z [ 1+(z_R/z)^{2}]} \, .
\end{equation}
Optimal detector positions are at the planes of maximal
wavefront curvature: $z_{\mathrm{opt}} = \pm z_R$, whereby the quantum
limit is saturated; i.e., $\mathcal{F}_{\mathrm{opt}} =
\mathcal{Q}$. In these planes, all the information about the
axial waist location is encoded in the intensity and can be extracted
with conventional imaging, thus avoiding more complicated and
less robust techniques.

Potential applications of this effect benefit from using a  relay
optical system for reimaging the object and obtaining a more
convenient detector position. Figure~\ref{figGauss} sketches the
simplest case of a thin lens placed a distance $z$ from the waist. Primed
symbols will distinguish henceforth parameters in the image space.  

\begin{figure}
  \includegraphics[width=0.85\columnwidth]{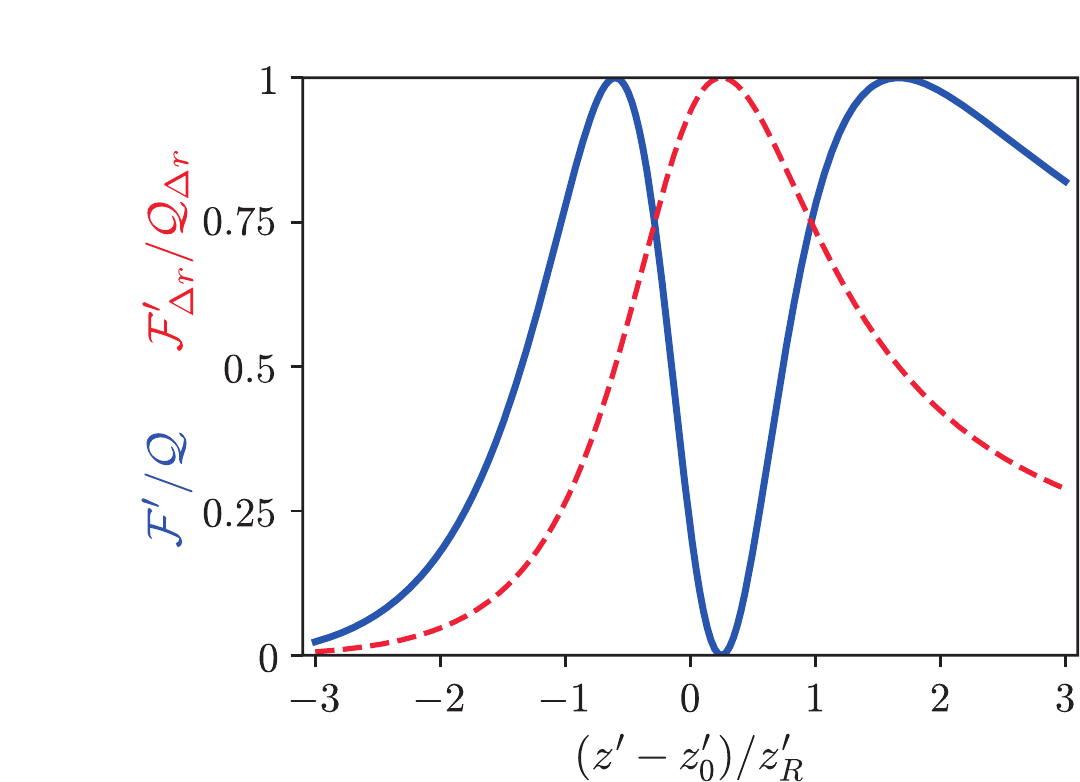}
  \caption{Fisher information in the image space in units of quantum
    Fisher information for different positions of the detector. We use 
     $z=5$, $f=1$, $z_R=1$, $w_0=1$ and the detector
    positions are relative to the beam waist in units of
    $z_R^{\prime}$.}
  \label{figFish}
\end{figure}

Since the ideal  imaging system applies a unitary transformation, the
qFI  does not change from the object space to the image space:
$\mathcal{Q}^{\prime}= \mathcal{Q}$.  
Recalling the standard relations~\cite{Siegman:1986aa}  
\begin{equation}
  w_{0}^{\prime 2} = m^2 w_0^2, \qquad
  z_R^{\prime} = m^2 z_R ,
  \qquad
  z_0^{\prime} = m^2 (z - f) + f,
\end{equation} 
between the original and new beam parameters,
where $m^2=f^2/[(z-f)^2+z_R^2]$ is the magnification,
we find the beam width at the detector position $z^{\prime}$  to be
\begin{equation}
  w^{\prime 2}(z^{\prime}) = w_0^{\prime 2} \left[1+\left (
      \frac{z^{\prime} -z_0^{\prime}}{z_R^{\prime}}\right)^2\right] \, .
\end{equation}
Much in the same way as in Eq.~\eqref{res1}, we have now 
\begin{align}
  \mathcal{F}^{\prime}(z) & = 
 \frac{\partial_{z} w^{\prime 2}(z^{\prime} )}{w^{\prime 2}(z^{\prime})}\nonumber  \\
 & =  \frac{4(f-z^{\prime})^2[ z z^{\prime}-f(z+z^{\prime})]^2}
{(z^2+z_R^2)z'^2-2f z^{\prime}(z^2+z_R^2+z z^{\prime})+f^2[(z+z^{\prime})^2+z_R^2] }
       . 
\end{align}
The typical behavior of $\mathcal{F}^{\prime}$ around the beam waist is shown
in Fig.~\ref{figFish}. We observe the presence of well-resolved maxima
and minima. While the beam in the image space is symmetric about the
waist, the response of the beam width to small changes of the true
distance $z$ is different inside and outside the waist, which makes
the FI assymetrical with respect to the image waist. These
extremal points are located at
\begin{equation}
\label{optimal_plane}
z^{\prime}_{\mathrm{opt}} = \left \lbrace 
\begin{array}{l}
  z_0^{\prime} + \alpha z_R^{\prime} \, ,\\
  \\
\displaystyle z_0^{\prime}-\frac{1}{\alpha}z_R^{\prime}\, ,
\end{array}
\right.
\end{equation}
where $\alpha=(f-z-z_R)/(f-z+z_R)$.  In the geometrical limit
$f-z \gg z_R$, we have $\alpha \simeq1$ and
$z^{\prime}_{\mathrm{opt}} \approx z_0^{\prime }\pm z_R^{\prime}$, so
the asymmetry disappears.  Interestingly, information about axial
displacements is zero in the plane of the geometrical image
$z^{\prime}=f z/(z-f) $, as the FI tends to zero therein. In this
sense, optimal axial localization (requiring considerable image blur)
and transverse localization (benefiting from sharpness) complement
each other. PSF engineering reaches a balance to resolve this issue
and provides a good three-dimensional resolution. However, these
methods always broaden the PSF, even more than our defocusing in
$z_{R}$. 

\begin{figure}
\includegraphics[width=0.90\columnwidth]{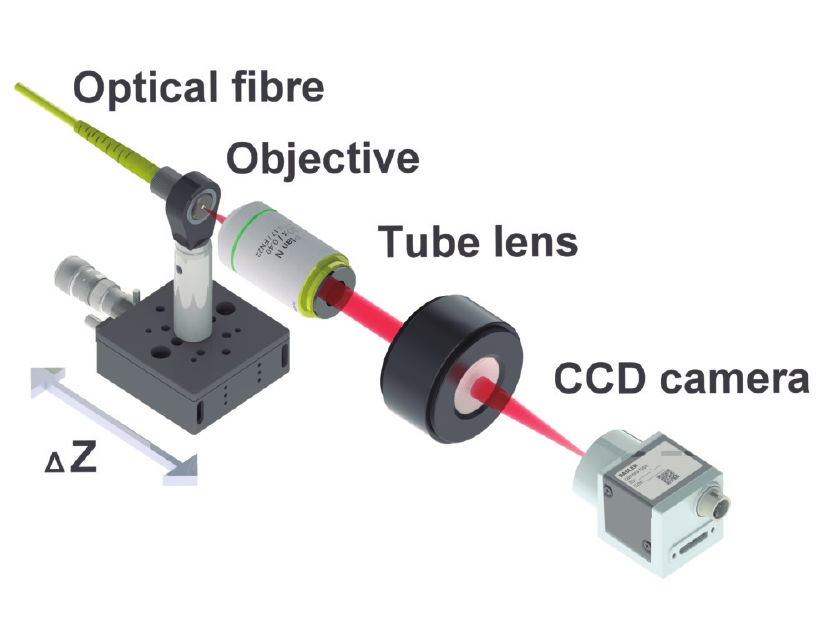}
\caption{Experimental setup used to measure the axial displacement
  $z$. See text for a detailed description.}
\label{figsetup} 
\end{figure}

\emph{Experiment.---}
To check the previous theory we have used a classical microscopy
setup, as schematized in Fig.~\ref{figsetup}. It consists of an
objective corrected for infinity and a tube lens, all together
providing a $20\times$ magnification of the output face of a single
mode fiber representing a Gaussian source. The fiber is coupled with a
632.8~nm He-Ne laser. As the Rayleigh range $z_{R}$ at the fiber
output is 18.9~$\mu$m, the camera with 5.5~$\mu$m is moved 7.6~mm out
of the system nominal image plane to become aligned with one of the
optimal detection positions given in
Eq.~\eqref{optimal_plane}. Controlled changes of the axial distance
$z$ were implemented by moving the fiber axially using a piezo stage
with a resolution of 1~nm.

\begin{figure}[t]
  \includegraphics[width=0.95\columnwidth]{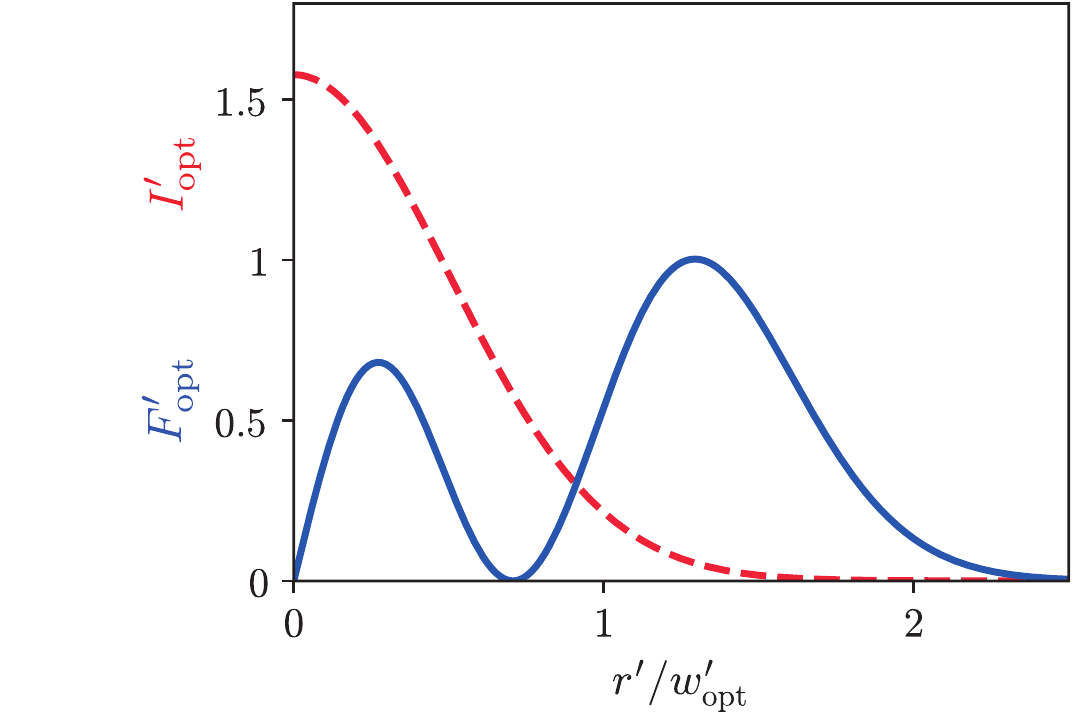}
  \caption{Normalized radial density of Fisher information (solid blue
    line) and  normalized beam  intensity profile (dashed orange line)
    in the optimal detection plane, where the beam has a width
    $w^{\prime}_{\mathrm{opt}}$. The system parameters are the same as
    in Fig.~\ref{figFish}.}
  \label{figdens}
\end{figure}

Note that the integrand in Eq.~\eqref{fishinf}; viz,
\begin{equation}
  F (r; z) = r\,   \frac{[ \partial_zp(r|z) ]^2}{p(r|z)}
\end{equation}
can be seen as a radial density of Fisher information. This magnitude
is plot in  Fig.~\ref{figdens} at the optimal detection plane, which
hints at constructing a robust estimator  of axial displacement from
the registered intensity scans. The information drops to zero at
$r_b=w_0^{\prime}/\sqrt{2}$  and the bulk
of information ($2/e \simeq 74\%$) resides outside this boundary in
the wings of the Gaussian intensity distribution. We will call
$I_{\mathrm{det}} (z)$ the intensity outside $r_{b}$ for the object distance 
$z$. Then for small displacements $\delta_z$ from the nominal position
$z$ we have
\begin{equation}
  \label{estimator}
  I_{\mathrm{det}} (z +\delta_z) =   I_{\mathrm{det}} (z)
  (1-\delta_z/z_R) \, .
\end{equation}
This linear relation is readily inverted to yield an
estimate $\widehat{\delta}_z$ of $\delta_z$ from $I_{\mathrm{det}}$.
Of course, we might be tempted to use the maximum likelihood estimator
based on the full profile. However, this estimator turns to be a bit
noisy due to systematic errors~\cite{Smith:2010aa}. On the contrary,
our estimator $\widehat{\delta}_{z}$ is simple and
robust. Nevertheless, we stress that we are interested in a
proof-of-concept experiment, so small deviations from the theoretical
best performance are non issue. 

We also notice that we are assuming that  the nominal axial distance
is known. This is not a serious drawback, as one can perform a
previous calibration (as we did in the experiment), and then measure
in a very precise manner around the nominal value. 

\begin{figure}[t]
  \includegraphics[width= \columnwidth]{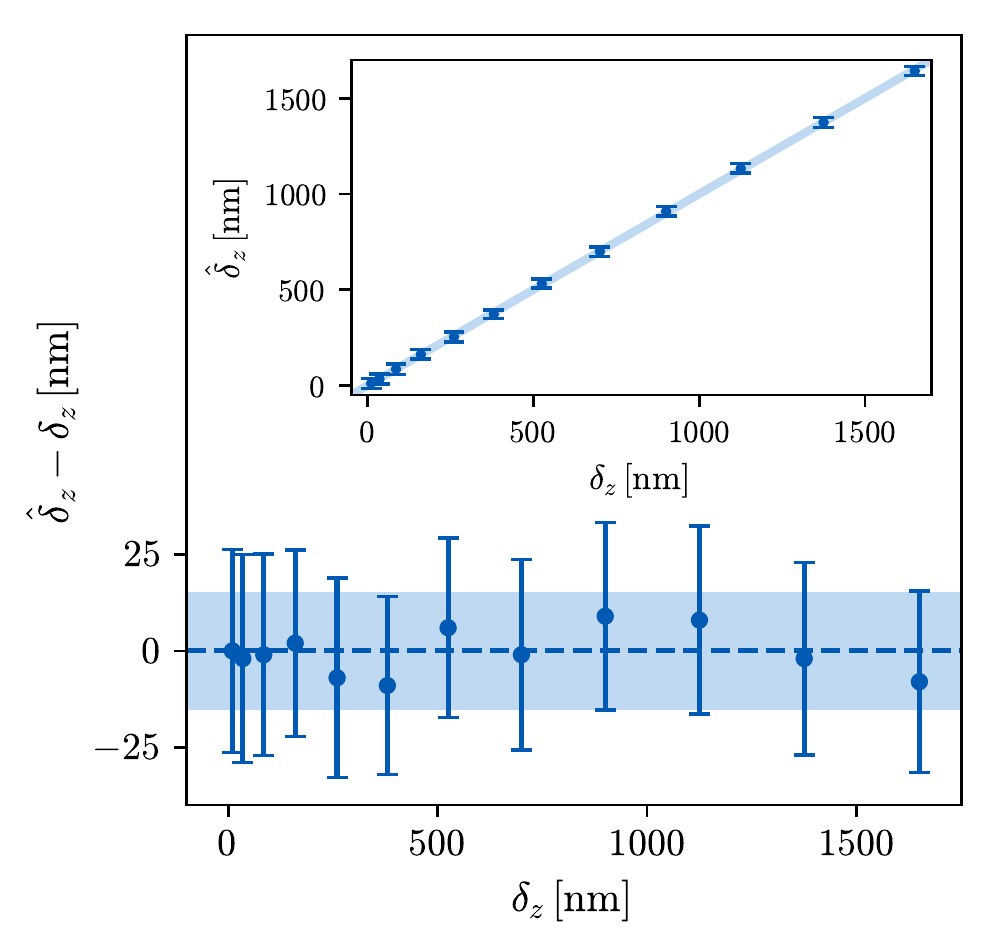}
  \caption{Experimental estimation of axial displacements $\delta_z$
    from the nominal object plane with respect to which the camera is
    optimally placed \eqref{optimal_plane}. The inset shows the
    statistics of the estimator $\widehat{\delta}_z$ as defined in
    Eq.~\eqref{estimator}. In the main plot he true distance was
    subtracted from the estimates to get a more convenient scale on
    the vertical axis. The blue strips depict the quantum bound for
    the $2\times 10^6$ detections and $z_R=18.9~\mu$m. \label{figExp}}
\end{figure}

Our experimental results are summarized in
Fig.~\ref{figExp}. Measurement errors are consistent across the full
range of axial displacements $\delta_z\in$[10~nm, 1650~nm] averaging $24.8$~nm. This
is about 800 times below the depth of focus $z_R$ and not much above
the quantum limit of $14.9$~nm corresponding to the total number of
$1.6\times 10^6$ detections registered for each $\delta_z$ setting.

Thus far  we  have focused on axial measurements with Gaussian
beams. What about \textit{uncooperative} point sources? In this case, the
source generates a spherical (paraboidal) wavefront, which after
transiting a distance $z$ enters an imaging system that truncates the
unbounded wave with a pupil function. Keeping things simple and
considering a Gaussian pupil transmisivity of width $w_l$ the wave on
the pupil depends on $z$ through
\begin{equation}
  U(x,y;z)=\sqrt{\frac{2}{\pi w_l^2}}
    \exp\left [ - \frac{r^2}{w_l^2}- i  \frac{kr^{2}}{2(z - f)} \right ]  \, .
\end{equation}
The state in the aperture is not just axial propagation from the
point source, as  the pupil acts like a filter and one needs to
renormalize the state. The process is now not unitary and the qFI
cannot be calculated in terms of a generator $G$. Instead, one has
\begin{equation}
\label{qfi}
\tfrac{1}{4} \mathcal{Q} (z) = \langle \partial_{z}\Psi(z)|
\partial_{z}\Psi(z) \rangle -
\langle \partial_{z}\Psi(z)| \Psi (z) \rangle\langle \Psi (z) |
\partial_{z}\Psi(z) \rangle \, .
\end{equation}
At difference of a Gaussian source, the result now reads
\begin{equation}
\mathcal{Q} (z) =\frac{k^2 w_l^4}{4 z^4} \, , 
\end{equation}
which strongly depends on the true distance $z$. We mention in passing
that, like for a Gaussian source,  this qFI  can be
saturated with a single intensity scan optimally placed with respect
to the nominal image plane. It is intriguing to note that
$n=2\times 10^6$ detections like in our experiment registered with a
one meter aperture $w_l=1$~m in visible light $k=10^7$m$^{-1}$ would
theoretically provide axial localization of a point source in a low
Earth orbit $z=200$~km with about 5~m accuracy.

In conclusion, we have theoretically and experimentally demonstrated
the axial superresolution based on direct detection. The quantum
limits can be saturated with a single intensity scan provided the
camera is placed in one of two optimal transversal detection planes.
Hence for axial localization problem there is no advantage in adopting
more complicated detection schemes. Our method makes three-dimensional
superresolution imaging promising and can be potentially useful for
enhancing the resolution of optical microscopes.

We thank Robert W. Boyd for helpful discussions. We acknowledge
financial support from the Grant Agency of the Czech Republic (Grant
No.~18-04291S), the Palack\'y University (Grant
No. IGA\_PrF\_2019\_007), and the Spanish MINECO (Grant
FIS2015-67963-P).


\begin{thebibliography}{32}%
\makeatletter
\providecommand \@ifxundefined [1]{%
 \@ifx{#1\undefined}
}%
\providecommand \@ifnum [1]{%
 \ifnum #1\expandafter \@firstoftwo
 \else \expandafter \@secondoftwo
 \fi
}%
\providecommand \@ifx [1]{%
 \ifx #1\expandafter \@firstoftwo
 \else \expandafter \@secondoftwo
 \fi
}%
\providecommand \natexlab [1]{#1}%
\providecommand \enquote  [1]{``#1''}%
\providecommand \bibnamefont  [1]{#1}%
\providecommand \bibfnamefont [1]{#1}%
\providecommand \citenamefont [1]{#1}%
\providecommand \href@noop [0]{\@secondoftwo}%
\providecommand \href [0]{\begingroup \@sanitize@url \@href}%
\providecommand \@href[1]{\@@startlink{#1}\@@href}%
\providecommand \@@href[1]{\endgroup#1\@@endlink}%
\providecommand \@sanitize@url [0]{\catcode `\\12\catcode `\$12\catcode
  `\&12\catcode `\#12\catcode `\^12\catcode `\_12\catcode `\%12\relax}%
\providecommand \@@startlink[1]{}%
\providecommand \@@endlink[0]{}%
\providecommand \url  [0]{\begingroup\@sanitize@url \@url }%
\providecommand \@url [1]{\endgroup\@href {#1}{\urlprefix }}%
\providecommand \urlprefix  [0]{URL }%
\providecommand \Eprint [0]{\href }%
\providecommand \doibase [0]{http://dx.doi.org/}%
\providecommand \selectlanguage [0]{\@gobble}%
\providecommand \bibinfo  [0]{\@secondoftwo}%
\providecommand \bibfield  [0]{\@secondoftwo}%
\providecommand \translation [1]{[#1]}%
\providecommand \BibitemOpen [0]{}%
\providecommand \bibitemStop [0]{}%
\providecommand \bibitemNoStop [0]{.\EOS\space}%
\providecommand \EOS [0]{\spacefactor3000\relax}%
\providecommand \BibitemShut  [1]{\csname bibitem#1\endcsname}%
\let\auto@bib@innerbib\@empty
\bibitem [{\citenamefont {Rayleigh}(1879)}]{Rayleigh:1879aa}%
  \BibitemOpen
  \bibfield  {author} {\bibinfo {author} {\bibfnamefont {Lord}\ \bibnamefont
  {Rayleigh}},\ }\bibfield  {title} {\enquote {\bibinfo {title} {Investigations
  in {O}ptics, with special reference to the spectroscope},}\ }\href@noop {}
  {\bibfield  {journal} {\bibinfo  {journal} {Phil. Mag.}\ }\textbf {\bibinfo
  {volume} {8}},\ \bibinfo {pages} {261--274, 403--411, 477--486} (\bibinfo
  {year} {1879})}\BibitemShut {NoStop}%
\bibitem [{\citenamefont {Abbe}(1873)}]{Abbe:1873aa}%
  \BibitemOpen
  \bibfield  {author} {\bibinfo {author} {\bibfnamefont {E.}~\bibnamefont
  {Abbe}},\ }\bibfield  {title} {\enquote {\bibinfo {title} {Ueber einen neuen
  {B}eleuchtungsapparat am {M}ikroskop},}\ }\href {\doibase 10.1007/BF02956177}
  {\bibfield  {journal} {\bibinfo  {journal} {Arch. Mikrosk. Anat.}\ }\textbf
  {\bibinfo {volume} {9}},\ \bibinfo {pages} {469--480} (\bibinfo {year}
  {1873})}\BibitemShut {NoStop}%
\bibitem [{\citenamefont {Ram}\ \emph {et~al.}(2006)\citenamefont {Ram},
  \citenamefont {Ward},\ and\ \citenamefont {Ober}}]{Ram:2006aa}%
  \BibitemOpen
  \bibfield  {author} {\bibinfo {author} {\bibfnamefont {S.}~\bibnamefont
  {Ram}}, \bibinfo {author} {\bibfnamefont {E.~S.}\ \bibnamefont {Ward}}, \
  and\ \bibinfo {author} {\bibfnamefont {R.~J.}\ \bibnamefont {Ober}},\
  }\bibfield  {title} {\enquote {\bibinfo {title} {Beyond {R}ayleigh's
  criterion: A resolution measure with application to single-molecule
  microscopy},}\ }\href {http://www.pnas.org/content/103/12/4457.abstract}
  {\bibfield  {journal} {\bibinfo  {journal} {PNAS}\ }\textbf {\bibinfo
  {volume} {103}},\ \bibinfo {pages} {4457--4462} (\bibinfo {year}
  {2006})}\BibitemShut {NoStop}%
\bibitem [{\citenamefont {Hell}(2007)}]{Hell:2007aa}%
  \BibitemOpen
  \bibfield  {author} {\bibinfo {author} {\bibfnamefont {S.~W.}\ \bibnamefont
  {Hell}},\ }\bibfield  {title} {\enquote {\bibinfo {title} {Far-field optical
  nanoscopy},}\ }\href
  {http://science.sciencemag.org/content/316/5828/1153.abstract} {\bibfield
  {journal} {\bibinfo  {journal} {Science}\ }\textbf {\bibinfo {volume}
  {316}},\ \bibinfo {pages} {1153--1158} (\bibinfo {year} {2007})}\BibitemShut
  {NoStop}%
\bibitem [{\citenamefont {Huang}\ \emph {et~al.}(2009)\citenamefont {Huang},
  \citenamefont {Bates},\ and\ \citenamefont {Zhuang}}]{Huang:2009aa}%
  \BibitemOpen
  \bibfield  {author} {\bibinfo {author} {\bibfnamefont {B.}~\bibnamefont
  {Huang}}, \bibinfo {author} {\bibfnamefont {M.}~\bibnamefont {Bates}}, \ and\
  \bibinfo {author} {\bibfnamefont {X.}~\bibnamefont {Zhuang}},\ }\bibfield
  {title} {\enquote {\bibinfo {title} {Super-resolution fluorescence
  microscopy},}\ }\href {\doibase 10.1146/annurev.biochem.77.061906.092014}
  {\bibfield  {journal} {\bibinfo  {journal} {Annu. Rev. Biochem.}\ }\textbf
  {\bibinfo {volume} {78}},\ \bibinfo {pages} {993--1016} (\bibinfo {year}
  {2009})}\BibitemShut {NoStop}%
\bibitem [{\citenamefont {Huszka}\ and\ \citenamefont
  {Gijs}(2019)}]{Huszka:2019aa}%
  \BibitemOpen
  \bibfield  {author} {\bibinfo {author} {\bibfnamefont {G.}~\bibnamefont
  {Huszka}}\ and\ \bibinfo {author} {\bibfnamefont {M.~A.~M.}\ \bibnamefont
  {Gijs}},\ }\bibfield  {title} {\enquote {\bibinfo {title} {Super-resolution
  optical imaging: A comparison},}\ }\href {\doibase
  https://doi.org/10.1016/j.mne.2018.11.005} {\bibfield  {journal} {\bibinfo
  {journal} {MNE}\ }\textbf {\bibinfo {volume} {2}},\ \bibinfo {pages} {7--28}
  (\bibinfo {year} {2019})}\BibitemShut {NoStop}%
\bibitem [{\citenamefont {Schermelleh}\ \emph {et~al.}(2010)\citenamefont
  {Schermelleh}, \citenamefont {Heintzmann},\ and\ \citenamefont
  {Leonhardt}}]{Schermelleh:2010aa}%
  \BibitemOpen
  \bibfield  {author} {\bibinfo {author} {\bibfnamefont {L.}~\bibnamefont
  {Schermelleh}}, \bibinfo {author} {\bibfnamefont {R.}~\bibnamefont
  {Heintzmann}}, \ and\ \bibinfo {author} {\bibfnamefont {H.}~\bibnamefont
  {Leonhardt}},\ }\bibfield  {title} {\enquote {\bibinfo {title} {A guide to
  super-resolution fluorescence microscopy},}\ }\href
  {http://jcb.rupress.org/content/190/2/165.abstract} {\bibfield  {journal}
  {\bibinfo  {journal} {J. Cell Biol.}\ }\textbf {\bibinfo {volume} {190}},\
  \bibinfo {pages} {165} (\bibinfo {year} {2010})}\BibitemShut {NoStop}%
\bibitem [{\citenamefont {Leung}\ and\ \citenamefont
  {Chou}(2011)}]{Leung:2011aa}%
  \BibitemOpen
  \bibfield  {author} {\bibinfo {author} {\bibfnamefont {B.~O.}\ \bibnamefont
  {Leung}}\ and\ \bibinfo {author} {\bibfnamefont {K.~C.}\ \bibnamefont
  {Chou}},\ }\bibfield  {title} {\enquote {\bibinfo {title} {Review of
  super-resolution fluorescence microscopy for biology},}\ }\href {\doibase
  10.1366/11-06398} {\bibfield  {journal} {\bibinfo  {journal} {Appl.
  Spectrosc.}\ }\textbf {\bibinfo {volume} {65}},\ \bibinfo {pages} {967--980}
  (\bibinfo {year} {2011})}\BibitemShut {NoStop}%
\bibitem [{\citenamefont {Hell}\ and\ \citenamefont
  {Wichmann}(1994)}]{Hell:1994ab}%
  \BibitemOpen
  \bibfield  {author} {\bibinfo {author} {\bibfnamefont {S.~W.}\ \bibnamefont
  {Hell}}\ and\ \bibinfo {author} {\bibfnamefont {J.}~\bibnamefont
  {Wichmann}},\ }\bibfield  {title} {\enquote {\bibinfo {title} {Breaking the
  diffraction resolution limit by stimulated emission:
  stimulated-emission-depletion fluorescence microscopy},}\ }\href {\doibase
  10.1364/OL.19.000780} {\bibfield  {journal} {\bibinfo  {journal} {Opt.
  Lett.}\ }\textbf {\bibinfo {volume} {19}},\ \bibinfo {pages} {780--782}
  (\bibinfo {year} {1994})}\BibitemShut {NoStop}%
\bibitem [{\citenamefont {Betzig}\ \emph {et~al.}(2006)\citenamefont {Betzig},
  \citenamefont {Patterson}, \citenamefont {Sougrat}, \citenamefont
  {Lindwasser}, \citenamefont {Olenych}, \citenamefont {Bonifacino},
  \citenamefont {Davidson}, \citenamefont {Lippincott-Schwartz},\ and\
  \citenamefont {Hess}}]{Betzig:2006aa}%
  \BibitemOpen
  \bibfield  {author} {\bibinfo {author} {\bibfnamefont {E.}~\bibnamefont
  {Betzig}}, \bibinfo {author} {\bibfnamefont {G.~H.}\ \bibnamefont
  {Patterson}}, \bibinfo {author} {\bibfnamefont {R.}~\bibnamefont {Sougrat}},
  \bibinfo {author} {\bibfnamefont {O.~W.}\ \bibnamefont {Lindwasser}},
  \bibinfo {author} {\bibfnamefont {S.}~\bibnamefont {Olenych}}, \bibinfo
  {author} {\bibfnamefont {J.~S.}\ \bibnamefont {Bonifacino}}, \bibinfo
  {author} {\bibfnamefont {M.~W.}\ \bibnamefont {Davidson}}, \bibinfo {author}
  {\bibfnamefont {J.}~\bibnamefont {Lippincott-Schwartz}}, \ and\ \bibinfo
  {author} {\bibfnamefont {H.~F.}\ \bibnamefont {Hess}},\ }\bibfield  {title}
  {\enquote {\bibinfo {title} {Imaging intracellular fluorescent proteins at
  nanometer resolution},}\ }\href
  {http://science.sciencemag.org/content/313/5793/1642.abstract} {\bibfield
  {journal} {\bibinfo  {journal} {Science}\ }\textbf {\bibinfo {volume}
  {313}},\ \bibinfo {pages} {1642} (\bibinfo {year} {2006})}\BibitemShut
  {NoStop}%
\bibitem [{\citenamefont {Huang}\ \emph {et~al.}(2008)\citenamefont {Huang},
  \citenamefont {Wang}, \citenamefont {Bates},\ and\ \citenamefont
  {Zhuang}}]{Huang:2008aa}%
  \BibitemOpen
  \bibfield  {author} {\bibinfo {author} {\bibfnamefont {B.}~\bibnamefont
  {Huang}}, \bibinfo {author} {\bibfnamefont {W.}~\bibnamefont {Wang}},
  \bibinfo {author} {\bibfnamefont {M.}~\bibnamefont {Bates}}, \ and\ \bibinfo
  {author} {\bibfnamefont {X.}~\bibnamefont {Zhuang}},\ }\bibfield  {title}
  {\enquote {\bibinfo {title} {Three-dimensional super-resolution imaging by
  stochastic optical reconstruction microscopy},}\ }\href {\doibase
  10.1126/science.1153529} {\bibfield  {journal} {\bibinfo  {journal}
  {Science}\ }\textbf {\bibinfo {volume} {319}},\ \bibinfo {pages} {810}
  (\bibinfo {year} {2008})}\BibitemShut {NoStop}%
\bibitem [{\citenamefont {Pavani}\ \emph {et~al.}(2009)\citenamefont {Pavani},
  \citenamefont {Thompson}, \citenamefont {Biteen}, \citenamefont {Lord},
  \citenamefont {Liu}, \citenamefont {Twieg}, \citenamefont {Piestun},\ and\
  \citenamefont {Moerner}}]{Pavani:2009aa}%
  \BibitemOpen
  \bibfield  {author} {\bibinfo {author} {\bibfnamefont {S.~R.~P.}\
  \bibnamefont {Pavani}}, \bibinfo {author} {\bibfnamefont {M.~A.}\
  \bibnamefont {Thompson}}, \bibinfo {author} {\bibfnamefont {J.~S.}\
  \bibnamefont {Biteen}}, \bibinfo {author} {\bibfnamefont {S.~J.}\
  \bibnamefont {Lord}}, \bibinfo {author} {\bibfnamefont {N.}~\bibnamefont
  {Liu}}, \bibinfo {author} {\bibfnamefont {R.~J.}\ \bibnamefont {Twieg}},
  \bibinfo {author} {\bibfnamefont {R.}~\bibnamefont {Piestun}}, \ and\
  \bibinfo {author} {\bibfnamefont {W.~E.}\ \bibnamefont {Moerner}},\
  }\bibfield  {title} {\enquote {\bibinfo {title} {Three-dimensional,
  single-molecule fluorescence imaging beyond the diffraction limit by using a
  double-helix point spread function},}\ }\href {\doibase
  10.1073/pnas.0900245106} {\bibfield  {journal} {\bibinfo  {journal} {Proc.
  Natl. Acad. Sci. USA}\ }\textbf {\bibinfo {volume} {106}},\ \bibinfo {pages}
  {2995} (\bibinfo {year} {2009})}\BibitemShut {NoStop}%
\bibitem [{\citenamefont {Jia}\ \emph {et~al.}(2014)\citenamefont {Jia},
  \citenamefont {Vaughan},\ and\ \citenamefont {Zhuang}}]{Jia:2014aa}%
  \BibitemOpen
  \bibfield  {author} {\bibinfo {author} {\bibfnamefont {S.}~\bibnamefont
  {Jia}}, \bibinfo {author} {\bibfnamefont {J.~C.}\ \bibnamefont {Vaughan}}, \
  and\ \bibinfo {author} {\bibfnamefont {X.}~\bibnamefont {Zhuang}},\
  }\bibfield  {title} {\enquote {\bibinfo {title} {Isotropic three-dimensional
  super-resolution imaging with a self-bending point spread function},}\ }\href
  {https://doi.org/10.1038/nphoton.2014.13} {\bibfield  {journal} {\bibinfo
  {journal} {Nat. Photonics}\ }\textbf {\bibinfo {volume} {8}},\ \bibinfo
  {pages} {302} (\bibinfo {year} {2014})}\BibitemShut {NoStop}%
\bibitem [{\citenamefont {Tamburini}\ \emph {et~al.}(2006)\citenamefont
  {Tamburini}, \citenamefont {Anzolin}, \citenamefont {Umbriaco}, \citenamefont
  {Bianchini},\ and\ \citenamefont {Barbieri}}]{Tamburini:2006aa}%
  \BibitemOpen
  \bibfield  {author} {\bibinfo {author} {\bibfnamefont {F.}~\bibnamefont
  {Tamburini}}, \bibinfo {author} {\bibfnamefont {G.}~\bibnamefont {Anzolin}},
  \bibinfo {author} {\bibfnamefont {G.}~\bibnamefont {Umbriaco}}, \bibinfo
  {author} {\bibfnamefont {A.}~\bibnamefont {Bianchini}}, \ and\ \bibinfo
  {author} {\bibfnamefont {C.}~\bibnamefont {Barbieri}},\ }\bibfield  {title}
  {\enquote {\bibinfo {title} {Overcoming the {R}ayleigh criterion limit with
  optical vortices},}\ }\href {\doibase 10.1103/PhysRevLett.97.163903}
  {\bibfield  {journal} {\bibinfo  {journal} {Phys. Rev. Lett.}\ }\textbf
  {\bibinfo {volume} {97}},\ \bibinfo {pages} {163903} (\bibinfo {year}
  {2006})}\BibitemShut {NoStop}%
\bibitem [{\citenamefont {Pa{\'u}r}\ \emph {et~al.}(2018)\citenamefont
  {Pa{\'u}r}, \citenamefont {Stoklasa}, \citenamefont {Grover}, \citenamefont
  {Krzic}, \citenamefont {S{\'a}nchez-Soto}, \citenamefont {Hradil},\ and\
  \citenamefont {{\v R}eh{\'a}{\v c}ek}}]{Paur:2018aa}%
  \BibitemOpen
  \bibfield  {author} {\bibinfo {author} {\bibfnamefont {M.}~\bibnamefont
  {Pa{\'u}r}}, \bibinfo {author} {\bibfnamefont {B.}~\bibnamefont {Stoklasa}},
  \bibinfo {author} {\bibfnamefont {J.}~\bibnamefont {Grover}}, \bibinfo
  {author} {\bibfnamefont {A.}~\bibnamefont {Krzic}}, \bibinfo {author}
  {\bibfnamefont {L.~L.}\ \bibnamefont {S{\'a}nchez-Soto}}, \bibinfo {author}
  {\bibfnamefont {Z.}~\bibnamefont {Hradil}}, \ and\ \bibinfo {author}
  {\bibfnamefont {J.}~\bibnamefont {{\v R}eh{\'a}{\v c}ek}},\ }\bibfield
  {title} {\enquote {\bibinfo {title} {Tempering {R}ayleigh's curse with {PSF}
  shaping},}\ }\bibfield  {booktitle} {\emph {\bibinfo {booktitle} {Optica}},\
  }\href {\doibase 10.1364/OPTICA.5.001177} {\bibfield  {journal} {\bibinfo
  {journal} {Optica}\ }\textbf {\bibinfo {volume} {5}},\ \bibinfo {pages}
  {1177--1180} (\bibinfo {year} {2018})}\BibitemShut {NoStop}%
\bibitem [{\citenamefont {Dalgarno}\ \emph {et~al.}(2010)\citenamefont
  {Dalgarno}, \citenamefont {Dalgarno}, \citenamefont {Putoud}, \citenamefont
  {Lambert}, \citenamefont {Paterson}, \citenamefont {Logan}, \citenamefont
  {Towers}, \citenamefont {Warburton},\ and\ \citenamefont
  {Greenaway}}]{Dalgarno:2010aa}%
  \BibitemOpen
  \bibfield  {author} {\bibinfo {author} {\bibfnamefont {P.~A.}\ \bibnamefont
  {Dalgarno}}, \bibinfo {author} {\bibfnamefont {H.~I.~C.}\ \bibnamefont
  {Dalgarno}}, \bibinfo {author} {\bibfnamefont {A.}~\bibnamefont {Putoud}},
  \bibinfo {author} {\bibfnamefont {R.}~\bibnamefont {Lambert}}, \bibinfo
  {author} {\bibfnamefont {L.}~\bibnamefont {Paterson}}, \bibinfo {author}
  {\bibfnamefont {D.~C.}\ \bibnamefont {Logan}}, \bibinfo {author}
  {\bibfnamefont {D.~P.}\ \bibnamefont {Towers}}, \bibinfo {author}
  {\bibfnamefont {R.~J.}\ \bibnamefont {Warburton}}, \ and\ \bibinfo {author}
  {\bibfnamefont {A.~H.}\ \bibnamefont {Greenaway}},\ }\bibfield  {title}
  {\enquote {\bibinfo {title} {Multiplane imaging and three dimensional
  nanoscale particle tracking in biological microscopy},}\ }\href {\doibase
  10.1364/OE.18.000877} {\bibfield  {journal} {\bibinfo  {journal} {Opt.
  Express}\ }\textbf {\bibinfo {volume} {18}},\ \bibinfo {pages} {877--884}
  (\bibinfo {year} {2010})}\BibitemShut {NoStop}%
\bibitem [{\citenamefont {Juette}\ \emph {et~al.}(2008)\citenamefont {Juette},
  \citenamefont {Gould}, \citenamefont {Lessard}, \citenamefont {Mlodzianoski},
  \citenamefont {Nagpure}, \citenamefont {Bennett}, \citenamefont {Hess},\ and\
  \citenamefont {Bewersdorf}}]{Juette:2008aa}%
  \BibitemOpen
  \bibfield  {author} {\bibinfo {author} {\bibfnamefont {M.~F.}\ \bibnamefont
  {Juette}}, \bibinfo {author} {\bibfnamefont {T.~J.}\ \bibnamefont {Gould}},
  \bibinfo {author} {\bibfnamefont {M.~D.}\ \bibnamefont {Lessard}}, \bibinfo
  {author} {\bibfnamefont {M.~J.}\ \bibnamefont {Mlodzianoski}}, \bibinfo
  {author} {\bibfnamefont {B.~S.}\ \bibnamefont {Nagpure}}, \bibinfo {author}
  {\bibfnamefont {B.~T.}\ \bibnamefont {Bennett}}, \bibinfo {author}
  {\bibfnamefont {S.~T.}\ \bibnamefont {Hess}}, \ and\ \bibinfo {author}
  {\bibfnamefont {J.}~\bibnamefont {Bewersdorf}},\ }\bibfield  {title}
  {\enquote {\bibinfo {title} {Three-dimensional sub--100 nm resolution
  fluorescence microscopy of thick samples},}\ }\href
  {https://doi.org/10.1038/nmeth.1211} {\bibfield  {journal} {\bibinfo
  {journal} {Nat. Methods}\ }\textbf {\bibinfo {volume} {5}},\ \bibinfo {pages}
  {527} (\bibinfo {year} {2008})}\BibitemShut {NoStop}%
\bibitem [{\citenamefont {Abrahamsson}\ \emph {et~al.}(2012)\citenamefont
  {Abrahamsson}, \citenamefont {Chen}, \citenamefont {Hajj}, \citenamefont
  {Stallinga}, \citenamefont {Katsov}, \citenamefont {Wisniewski},
  \citenamefont {Mizuguchi}, \citenamefont {Soule}, \citenamefont {Mueller},
  \citenamefont {Darzacq}, \citenamefont {Darzacq}, \citenamefont {Wu},
  \citenamefont {Bargmann}, \citenamefont {Agard}, \citenamefont {Dahan},\ and\
  \citenamefont {Gustafsson}}]{Abrahamsson:2012aa}%
  \BibitemOpen
  \bibfield  {author} {\bibinfo {author} {\bibfnamefont {S.}~\bibnamefont
  {Abrahamsson}}, \bibinfo {author} {\bibfnamefont {J.}~\bibnamefont {Chen}},
  \bibinfo {author} {\bibfnamefont {B.}~\bibnamefont {Hajj}}, \bibinfo {author}
  {\bibfnamefont {S.}~\bibnamefont {Stallinga}}, \bibinfo {author}
  {\bibfnamefont {A.~Y.}\ \bibnamefont {Katsov}}, \bibinfo {author}
  {\bibfnamefont {J.}~\bibnamefont {Wisniewski}}, \bibinfo {author}
  {\bibfnamefont {G.}~\bibnamefont {Mizuguchi}}, \bibinfo {author}
  {\bibfnamefont {P.}~\bibnamefont {Soule}}, \bibinfo {author} {\bibfnamefont
  {F.}~\bibnamefont {Mueller}}, \bibinfo {author} {\bibfnamefont {C.~D.}\
  \bibnamefont {Darzacq}}, \bibinfo {author} {\bibfnamefont {X.}~\bibnamefont
  {Darzacq}}, \bibinfo {author} {\bibfnamefont {C.}~\bibnamefont {Wu}},
  \bibinfo {author} {\bibfnamefont {C.~I.}\ \bibnamefont {Bargmann}}, \bibinfo
  {author} {\bibfnamefont {D.~A.}\ \bibnamefont {Agard}}, \bibinfo {author}
  {\bibfnamefont {M.}~\bibnamefont {Dahan}}, \ and\ \bibinfo {author}
  {\bibfnamefont {M.~G.~L.}\ \bibnamefont {Gustafsson}},\ }\bibfield  {title}
  {\enquote {\bibinfo {title} {Fast multicolor {3D} imaging using
  aberration-corrected multifocus microscopy},}\ }\href
  {https://doi.org/10.1038/nmeth.2277} {\bibfield  {journal} {\bibinfo
  {journal} {Nat. Methods}\ }\textbf {\bibinfo {volume} {10}},\ \bibinfo
  {pages} {60} (\bibinfo {year} {2012})}\BibitemShut {NoStop}%
\bibitem [{\citenamefont {von Diezmann}\ \emph {et~al.}(2017)\citenamefont {von
  Diezmann}, \citenamefont {Shechtman},\ and\ \citenamefont
  {Moerner}}]{Diezmann:2017aa}%
  \BibitemOpen
  \bibfield  {author} {\bibinfo {author} {\bibfnamefont {A.}~\bibnamefont {von
  Diezmann}}, \bibinfo {author} {\bibfnamefont {Y.}~\bibnamefont {Shechtman}},
  \ and\ \bibinfo {author} {\bibfnamefont {W.~E.}\ \bibnamefont {Moerner}},\
  }\bibfield  {title} {\enquote {\bibinfo {title} {Three-dimensional
  localization of single molecules for super-resolution imaging and
  single-particle tracking},}\ }\href {\doibase 10.1021/acs.chemrev.6b00629}
  {\bibfield  {journal} {\bibinfo  {journal} {Chem. Rev.}\ }\textbf {\bibinfo
  {volume} {117}},\ \bibinfo {pages} {7244--7275} (\bibinfo {year}
  {2017})}\BibitemShut {NoStop}%
\bibitem [{\citenamefont {Tsang}(2015)}]{Tsang:2015aa}%
  \BibitemOpen
  \bibfield  {author} {\bibinfo {author} {\bibfnamefont {M.}~\bibnamefont
  {Tsang}},\ }\bibfield  {title} {\enquote {\bibinfo {title} {Quantum limits to
  optical point-source localization},}\ }\href {\doibase
  10.1364/OPTICA.2.000646} {\bibfield  {journal} {\bibinfo  {journal} {Optica}\
  }\textbf {\bibinfo {volume} {2}},\ \bibinfo {pages} {646--653} (\bibinfo
  {year} {2015})}\BibitemShut {NoStop}%
\bibitem [{\citenamefont {Backlund}\ \emph {et~al.}(2018)\citenamefont
  {Backlund}, \citenamefont {Shechtman},\ and\ \citenamefont
  {Walsworth}}]{Backlund:2018aa}%
  \BibitemOpen
  \bibfield  {author} {\bibinfo {author} {\bibfnamefont {M.~P.}\ \bibnamefont
  {Backlund}}, \bibinfo {author} {\bibfnamefont {Y.}~\bibnamefont {Shechtman}},
  \ and\ \bibinfo {author} {\bibfnamefont {R.~L.}\ \bibnamefont {Walsworth}},\
  }\bibfield  {title} {\enquote {\bibinfo {title} {Fundamental precision bounds
  for three-dimensional optical localization microscopy with {P}oisson
  statistics},}\ }\href {\doibase 10.1103/PhysRevLett.121.023904} {\bibfield
  {journal} {\bibinfo  {journal} {Phys. Rev. Lett.}\ }\textbf {\bibinfo
  {volume} {121}},\ \bibinfo {pages} {023904} (\bibinfo {year}
  {2018})}\BibitemShut {NoStop}%
\bibitem [{\citenamefont {Petz}\ and\ \citenamefont
  {Ghinea}(2011)}]{Petz:2011aa}%
  \BibitemOpen
  \bibfield  {author} {\bibinfo {author} {\bibfnamefont {D.}~\bibnamefont
  {Petz}}\ and\ \bibinfo {author} {\bibfnamefont {C.}~\bibnamefont {Ghinea}},\
  }\enquote {\bibinfo {title} {Introduction to {Q}uantum {F}isher
  {I}nformation},}\ in\ \href {\doibase doi:10.1142/9789814338745_0015} {\emph
  {\bibinfo {booktitle} {Quantum Probability and Related Topics}}},\ Vol.\
  \bibinfo {volume} {Volume 27}\ (\bibinfo  {publisher} {World Scientific},\
  \bibinfo {year} {2011})\ pp.\ \bibinfo {pages} {261--281}\BibitemShut
  {NoStop}%
\bibitem [{\citenamefont {Tsang}\ \emph {et~al.}(2016)\citenamefont {Tsang},
  \citenamefont {Nair},\ and\ \citenamefont {Lu}}]{Tsang:2016aa}%
  \BibitemOpen
  \bibfield  {author} {\bibinfo {author} {\bibfnamefont {M.}~\bibnamefont
  {Tsang}}, \bibinfo {author} {\bibfnamefont {R.}~\bibnamefont {Nair}}, \ and\
  \bibinfo {author} {\bibfnamefont {X.-M.}\ \bibnamefont {Lu}},\ }\bibfield
  {title} {\enquote {\bibinfo {title} {Quantum theory of superresolution for
  two incoherent optical point sources},}\ }\href
  {http://link.aps.org/doi/10.1103/PhysRevX.6.031033} {\bibfield  {journal}
  {\bibinfo  {journal} {Phys. Rev. X}\ }\textbf {\bibinfo {volume} {6}},\
  \bibinfo {pages} {031033} (\bibinfo {year} {2016})}\BibitemShut {NoStop}%
\bibitem [{\citenamefont {Nair}\ and\ \citenamefont
  {Tsang}(2016)}]{Nair:2016aa}%
  \BibitemOpen
  \bibfield  {author} {\bibinfo {author} {\bibfnamefont {R.}~\bibnamefont
  {Nair}}\ and\ \bibinfo {author} {\bibfnamefont {M.}~\bibnamefont {Tsang}},\
  }\bibfield  {title} {\enquote {\bibinfo {title} {Far-field superresolution of
  thermal electromagnetic sources at the quantum limit},}\ }\href
  {http://arxiv.org/abs/1604.00937} {\bibfield  {journal} {\bibinfo  {journal}
  {Phys. Rev. Lett.}\ }\textbf {\bibinfo {volume} {117}},\ \bibinfo {pages}
  {190801} (\bibinfo {year} {2016})}\BibitemShut {NoStop}%
\bibitem [{\citenamefont {Ang}\ \emph {et~al.}(2016)\citenamefont {Ang},
  \citenamefont {Nair},\ and\ \citenamefont {Tsang}}]{Ang:2016aa}%
  \BibitemOpen
  \bibfield  {author} {\bibinfo {author} {\bibfnamefont {S.~Z.}\ \bibnamefont
  {Ang}}, \bibinfo {author} {\bibfnamefont {R.}~\bibnamefont {Nair}}, \ and\
  \bibinfo {author} {\bibfnamefont {M.}~\bibnamefont {Tsang}},\ }\bibfield
  {title} {\enquote {\bibinfo {title} {Quantum limit for two-dimensional
  resolution of two incoherent optical point sources},}\ }\href
  {http://arxiv.org/pdf/1606.00603.pdf} {\bibfield  {journal} {\bibinfo
  {journal} {Phys. Rev. A}\ }\textbf {\bibinfo {volume} {95}},\ \bibinfo
  {pages} {063847} (\bibinfo {year} {2016})}\BibitemShut {NoStop}%
\bibitem [{\citenamefont {Tsang}(2017)}]{Tsang:2017aa}%
  \BibitemOpen
  \bibfield  {author} {\bibinfo {author} {\bibfnamefont {M.}~\bibnamefont
  {Tsang}},\ }\bibfield  {title} {\enquote {\bibinfo {title} {Subdiffraction
  incoherent optical imaging via spatial-mode demultiplexing},}\ }\href
  {http://stacks.iop.org/1367-2630/19/i=2/a=023054} {\bibfield  {journal}
  {\bibinfo  {journal} {New J. Phys.}\ }\textbf {\bibinfo {volume} {19}},\
  \bibinfo {pages} {023054} (\bibinfo {year} {2017})}\BibitemShut {NoStop}%
\bibitem [{\citenamefont {Helstrom}(1976)}]{Helstrom:1976ij}%
  \BibitemOpen
  \bibfield  {author} {\bibinfo {author} {\bibfnamefont {C.~W.}\ \bibnamefont
  {Helstrom}},\ }\href@noop {} {\emph {\bibinfo {title} {Quantum {D}etection
  and {E}stimation {T}heory}}}\ (\bibinfo  {publisher} {Academic},\ \bibinfo
  {address} {New York},\ \bibinfo {year} {1976})\BibitemShut {NoStop}%
\bibitem [{\citenamefont {Holevo}(2003)}]{Holevo:2003fv}%
  \BibitemOpen
  \bibfield  {author} {\bibinfo {author} {\bibfnamefont {A.~S.}\ \bibnamefont
  {Holevo}},\ }\href@noop {} {\emph {\bibinfo {title} {Probabilistic and
  {S}tatistical {A}spects of {Q}uantum {T}heory}}},\ \bibinfo {edition} {2nd}\
  ed.\ (\bibinfo  {publisher} {North Holland},\ \bibinfo {address}
  {Amsterdam},\ \bibinfo {year} {2003})\BibitemShut {NoStop}%
\bibitem [{\citenamefont {Zhou}\ \emph {et~al.}(2019)\citenamefont {Zhou},
  \citenamefont {Yang}, \citenamefont {Hassett}, \citenamefont {Rafsanjani},
  \citenamefont {Mirhosseini}, \citenamefont {Vamivakas}, \citenamefont
  {Jordan}, \citenamefont {Shi},\ and\ \citenamefont {Boyd}}]{Zhou:2019aa}%
  \BibitemOpen
  \bibfield  {author} {\bibinfo {author} {\bibfnamefont {Y.}~\bibnamefont
  {Zhou}}, \bibinfo {author} {\bibfnamefont {J.}~\bibnamefont {Yang}}, \bibinfo
  {author} {\bibfnamefont {J.~D.}\ \bibnamefont {Hassett}}, \bibinfo {author}
  {\bibfnamefont {S.~M.~H.}\ \bibnamefont {Rafsanjani}}, \bibinfo {author}
  {\bibfnamefont {M.}~\bibnamefont {Mirhosseini}}, \bibinfo {author}
  {\bibfnamefont {A.~N.}\ \bibnamefont {Vamivakas}}, \bibinfo {author}
  {\bibfnamefont {A.~N.}\ \bibnamefont {Jordan}}, \bibinfo {author}
  {\bibfnamefont {Z.}~\bibnamefont {Shi}}, \ and\ \bibinfo {author}
  {\bibfnamefont {R.~W.}\ \bibnamefont {Boyd}},\ }\bibfield  {title} {\enquote
  {\bibinfo {title} {Quantum-limited estimation of the axial separation of two
  incoherent point sources},}\ }\href {\doibase 10.1364/OPTICA.6.000534}
  {\bibfield  {journal} {\bibinfo  {journal} {Optica}\ }\textbf {\bibinfo
  {volume} {6}},\ \bibinfo {pages} {534--541} (\bibinfo {year}
  {2019})}\BibitemShut {NoStop}%
\bibitem [{\citenamefont {Fuhrmann}\ \emph {et~al.}(2004)\citenamefont
  {Fuhrmann}, \citenamefont {Preza}, \citenamefont {O'Sullivan}, \citenamefont
  {Snyder},\ and\ \citenamefont {Smith}}]{Fuhrmann:2004aa}%
  \BibitemOpen
  \bibfield  {author} {\bibinfo {author} {\bibfnamefont {D.~R.}\ \bibnamefont
  {Fuhrmann}}, \bibinfo {author} {\bibfnamefont {C.}~\bibnamefont {Preza}},
  \bibinfo {author} {\bibfnamefont {J.~A.}\ \bibnamefont {O'Sullivan}},
  \bibinfo {author} {\bibfnamefont {D.~L.}\ \bibnamefont {Snyder}}, \ and\
  \bibinfo {author} {\bibfnamefont {W.~H.}\ \bibnamefont {Smith}},\ }\bibfield
  {title} {\enquote {\bibinfo {title} {Spectrum estimation from quantum-limited
  interferograms},}\ }\href {\doibase 10.1109/TSP.2004.824216} {\bibfield
  {journal} {\bibinfo  {journal} {IEEE Trans. Signal Process.}\ }\textbf
  {\bibinfo {volume} {52}},\ \bibinfo {pages} {950--961} (\bibinfo {year}
  {2004})}\BibitemShut {NoStop}%
\bibitem [{\citenamefont {Siegman}(1986)}]{Siegman:1986aa}%
  \BibitemOpen
  \bibfield  {author} {\bibinfo {author} {\bibfnamefont {A.E.}\ \bibnamefont
  {Siegman}},\ }\href@noop {} {\emph {\bibinfo {title} {Lasers}}}\ (\bibinfo
  {publisher} {Oxford University Press},\ \bibinfo {address} {Oxford},\
  \bibinfo {year} {1986})\BibitemShut {NoStop}%
\bibitem [{\citenamefont {Smith}\ \emph {et~al.}(2010)\citenamefont {Smith},
  \citenamefont {Joseph}, \citenamefont {Rieger},\ and\ \citenamefont
  {Lidke}}]{Smith:2010aa}%
  \BibitemOpen
  \bibfield  {author} {\bibinfo {author} {\bibfnamefont {C.~S}\ \bibnamefont
  {Smith}}, \bibinfo {author} {\bibfnamefont {N.}~\bibnamefont {Joseph}},
  \bibinfo {author} {\bibfnamefont {B.}~\bibnamefont {Rieger}}, \ and\ \bibinfo
  {author} {\bibfnamefont {K.~A.}\ \bibnamefont {Lidke}},\ }\bibfield  {title}
  {\enquote {\bibinfo {title} {Fast, single-molecule localization that achieves
  theoretically minimum uncertainty},}\ }\href
  {https://doi.org/10.1038/nmeth.1449} {\bibfield  {journal} {\bibinfo
  {journal} {Nat. Methods}\ }\textbf {\bibinfo {volume} {7}},\ \bibinfo {pages}
  {373} (\bibinfo {year} {2010})}\BibitemShut {NoStop}%
\end{thebibliography}

%

\end{document}